\documentclass[sn-mathphys-num]{sn-jnl}

\usepackage{graphicx}%
\usepackage{multirow}%
\usepackage{amsmath,amssymb,amsfonts}%
\usepackage{amsthm}%
\usepackage{mathrsfs}%
\usepackage[title]{appendix}%
\usepackage{xcolor}%
\usepackage{textcomp}%
\usepackage{manyfoot}%
\usepackage{booktabs}%
\usepackage{algorithm}%
\usepackage{algorithmicx}%
\usepackage{algpseudocode}%
\usepackage{listings}%

\theoremstyle{thmstyleone}%
\theoremstyle{thmstyletwo}%

\theoremstyle{thmstylethree}%

\begin{document}

\title[Approximate magic symmetry in a cobimaximal scenario of Majorana neutrinos]{Approximate magic symmetry in a cobimaximal scenario of Majorana neutrinos}

\author[1]{\fnm{Diana C.} \sur{Rivera-Agudelo}}\email{diana.rivera11@usc.edu.co}
\author*[1]{\fnm{S. L.} \sur{Tostado}}\email{sergio.tostado00@usc.edu.co}
\author[2]{\fnm{Nelson E.} \sur{Valderrama-Florez}}\email{nelson.valderrama2@unipamplona.edu.co}

\affil*[1]{\orgname{Universidad Santiago de Cali, Facultad de Ciencias B\'asicas, Campus Pampalinda}, \orgaddress{\street{Calle 5 No. 62-00}, \postcode{C\'odigo Postal 76001}, \state{Santiago de Cali}, \country{Colombia}}}

\affil[2]{\orgname{Universidad de Pamplona}, \orgaddress{km 1, v\'ia salida a Bucaramanga, Campus Universitario, \postcode{543050}, \state{Pamplona}, \country{Colombia}}}

%%%%%%%%%%%%%%%%% Abstract 
\abstract{
The searches for an underlying pattern in neutrino masses have motivated different proposals for textures in the neutrino mass matrix, which is also related to particular arrangements of the mixing matrix. Current experimental determinations of neutrino mixings have restricted some of the most studied proposals. In this work, we propose a new combined constraint on the neutrino mass matrix arising from the cobimaximal mixing and the magic symmetry in a model-independent fashion. We show that both conditions cannot be fulfilled simultaneously in an exact but rather in an approximate way. We have found that the Majorana CP phases are tightly restricted within this approach, leading to well-defined regions in the neutrinoless double beta decay amplitude. The small symmetry breaking requirement allows us to determine the Majorana phases for specific values of the lightest neutrino mass, consistent with current bounds on the mixing angles, neutrinoless double-beta decay, and cosmology results.
}


\maketitle
%
% ********** Text entry area ***************

\section{Introduction}

\label{sec:intro}

The mixing of massive neutrinos is summarized by the Pontecorvo-Maki-Nakagawa-Sakata (PMNS) matrix \cite{Pontecorvo:1957cp,Maki:1962mu}. The mixing angles have been long searched over the years and are actually known with high accuracy. Also, many efforts are being placed to confirm a nonzero $CP$ phase \cite{ParticleDataGroup:2022pth,Capozzi:2018ubv,deSalas:2020pgw,Esteban:2020cvm,Gonzalez-Garcia:2021dve,Abi:2018dnh}, but there is no compelling evidence for the Majorana phases as they are not involved in oscillation experiments. Nonetheless, the neutrinoless double beta decay ($0\nu\beta\beta$) experiments could be sensitive to nonzero Majorana phases \cite{Dolinski:2019nrj}, which would help elucidate the neutrino mass spectrum.

The PMNS matrix can be written in the standard way as
\begin{eqnarray}
\label{PMNS}
U_{\mathrm{PMNS}} &=& \left(
\begin{array}{c@{\quad}c@{\quad}c}
c_{12} c_{13} & s_{12}c_{13} & s_{13} e^{-i\delta_{CP}}
\\
- s_{12} c_{23} + c_{12}s_{23}s_{13}e^{i\delta_{CP}} & c_{12}c_{23} +
s_{12}s_{23}s_{13}e^{i\delta_{CP}} & -s_{23}c_{13}
\\
-s_{12}s_{23} - c_{12}c_{23}s_{13}e^{i\delta_{CP}} & c_{12}s_{23}
-c_{23}s_{12}s_{13}e^{i
\delta_{CP}} & c_{23}c_{13}
\end{array}
\right)
\nonumber
\\
&&\times{\mathrm{diag}} \left[e^{i\beta_{1}}, e^{i
\beta_{2}},1\right]~,
\end{eqnarray}
when Majorana neutrinos are taken into account. Here, $s_{ij}$ and $c_{ij}$ stand for
$\sin\theta_{ij}$ and $\cos\theta_{ij}$, respectively, with
$\theta_{ij}$ denoting the mixing angles $\theta_{12}$,
$\theta_{13}$, and $\theta_{23}$. The $CP$ violating phases are then written
as $\delta_{CP}$, $\beta_{1}$ and $\beta_{2}$, for Dirac and Majorana
phases, respectively.

A cobimaximal (CBM) mixing is recovered when values of $\theta_{23} = \pi/4$ and $\delta_{CP} = \pm\pi/2$ are taken in Eq.~(\ref{PMNS}) \cite{Fukuura:1999ze,Miura:2000sx,Harrison:2002et,Ma:2015fpa}, ignoring the Majorana phases. Although both values of the Dirac $CP$ phase lead to a cobimaximal matrix, the selection of $-\pi/2$ would be motivated by current experimental results within the $3\sigma$ interval of best-fit values.  The mixing matrix can be written as
\begin{equation}
\label{CBM}
U_{CBM} = \left(
\begin{array}{c@{\quad}c@{\quad}c}
c_{12}~c_{13} & s_{12}~c_{13} & i~s_{13}
\\
\frac{-1}{\sqrt{2}}(s_{12} + i~c_{12} ~s_{13}) & \frac{1}{\sqrt{2}}(c_{12}
- i~s_{12}~s_{13}) & \frac{- c_{13}}{\sqrt{2}}
\\
\frac{-1}{\sqrt{2}}(s_{12} - i~c_{12}~s_{13}) & \frac{1}{\sqrt{2}}(c_{12}
+ i~s_{12}~s_{13}) & \frac{c_{13}}{\sqrt{2}}
\end{array}
\right).
\end{equation}
While the atmospheric angle ($\theta_{23}$) and the Dirac phase are fixed to what is known as the maximal values, $\pi/4$ and $-\pi/2$, respectively, the solar and reactor angles remain free parameters, allowing them to be adjusted to match experimental results without altering the pattern in Eq. (\ref{CBM}). It is evident from this expression that  $|U_{\mu i}| = |U_{\tau i}|$, which is known as a $\mu-\tau$ symmetry. Additionally, the neutrino mass matrix obtained from $U_{CBM}$ \cite{Ma:2017moj}
\begin{equation}
\label{eq:massmatrix}
M_{\nu}^{0} = U_{CBM}{\mathrm{diag}}(m_{1},m_{2},m_{3}) U_{CBM}^{
	\mathrm{T}} = \left(
\begin{array}{c@{\quad}c@{\quad}c}
m_{ee} & m_{e\mu} & m_{e\mu}^{*}
\\
m_{e\mu} & m_{\mu\mu} & m_{\mu\tau}
\\
m_{e\mu}^{*}& m_{\mu\tau} & m_{\mu\mu}^{*}
\end{array}
\right),
\end{equation}
presents a reflection symmetry between the $\mu$ and $\tau$ labels. Such a texture of the neutrino mass matrix may be of interest in searches for improved models of neutrino masses and mixings since they could reflect the presence of additional symmetries\footnote{One possibility is to consider a discrete symmetry between neutrino states, type $A_{4}$, present at high energies, which should be broken at small energies to match the experimental data \cite{Ma:2015fpa}.} between specific flavor of neutrinos. One example is the partially flavor changing $CP$ transformations
\begin{equation}
\nu_{eL} \rightarrow(\nu_{eL})^{c} ,~~~~~ \nu_{\mu L}
\rightarrow(\nu_{\tau L})^{c} ,~~~~~ \nu_{\tau L} \rightarrow(
\nu_{\mu L})^{c}~,
\end{equation}
where $(\nu_{\alpha L})^{c}$ stands for the charge conjugate of $\nu_{\alpha L}$. Nevertheless, previous works have shown that by including Majorana phases in a CBM mixing matrix for writing the PMNS as $U_{CBM} \times{\mathrm{diag}} \left[ 1, e^{-i\frac{\alpha_{1}}{2}}, e^{-i\frac{\alpha_{2}}{2}}\right]$ the symmetry may break over \cite{Zhao:2017yvw,Gomez-Izquierdo:2023mph,Rivera-Agudelo:2022qpa}. In consequence, these phases could be used to parametrize the symmetry breaking. It is expected that the presence of additional symmetries in the neutrino mass matrix may guide the size of mixing angles, but also the amount of $CP$ violation. 

In this work, we investigate the presence of an additional magic symmetry in a neutrino mass matrix with a broken $\mu - \tau$ reflection symmetry through the Majorana phases, which, to the best of our knowledge, has not been previously explored. We discuss in section \ref{sec:magic} the imprints of a magic symmetry and its connection to the mixings. In section \ref{sec:deviations}, we study, analytically, the symmetry breaking by defining six parameters that account for deviations from the symmetric pattern, and examine their relation to neutrino mass hierarchies and Majorana phases. In section \ref{sec:results}, we present the numerical analysis of the symmetry breaking and the allowed regions of Majorana phases, which are also confronted with the neutrinoless double beta decay amplitude. Finally, in section \ref{sec:conclusions} we present our concluding remarks.

%s2 #&#
\section{Magic neutrino mass matrix}
%%LEAP%%%\label{sec2}
\label{sec:magic}

A $n\times n$ magic matrix $M$ is one where the sums of each column and row are identical $\sum_i^{n} M_{ij} =\sum_j^{n} M_{ij}$ \cite{Lam:2006wy}. A Majorana neutrino mass matrix $M_\nu$ is symmetric under the interchange $i\leftrightarrow j$, i.e. $[M_\nu]_{ij}=[M_\nu]_{ji}$, which has been the starting point for searches of specific textures of the mass matrix which give place to a magic sum \cite{Channey:2018cfj,Ouahid:2023hzs}. It has been found that a magic $M_{\nu}$ can be generated in the context of discrete flavor symmetries $A_4$ or $\Delta (54)$ within the type-I and type-II seesaw framework \cite{Verma:2019uiu,Singh:2022nmk,Ding:2024ozt,Altmannshofer:2024jyv}. 

The magic symmetry has been widely studied since naturally leads to a $TM_2$ pattern, where the second column of the PMNS matrix is constant \cite{Lam:2006wy}. This modification of the TBM pattern\footnote{Modifications of the TBM pattern are needed to generate a reactor angle different from zero and include $CP$ violation \cite{Rivera2,Rivera-Agudelo:2019seg}.} allows all the neutrino mixings be related to two parameters, giving place to some well-stablished correlations between the mixings, where one can identify the particular correlation between the reactor and solar angles  $\sin^2\theta_{12}=(3\cos^2 \theta_{13})^{-1}$. However, current global fits of $\theta_{13}$ lead to a very restricted range for $\theta_{12}$, which is in tension with the $3\sigma$ experimental range. Motivated by this, one could search for alternative mixing patterns compatible with an approximate magic symmetry.

Precise determinations of neutrino mixing parameters could be used to test the size of deviations from different symmetric patterns to explore the possibility of having an approximate rather than an exact symmetry at low energies. One could expect that these additional symmetries may be present at high energies, as described by any improved flavor model, where the number of free parameters would be strongly reduced given the restrictions of simultaneous symmetric scenarios. At small energies, one would have masses and mixing patterns different from the symmetric ones, but such deviations could be small. Then, one could search for such imprints given the expected precision of forthcoming experiments. 

 We should expect that demanding a small deviation from the symmetric scenarios may help restrict the parameters of the PMNS matrix and particular textures of the neutrino mass matrix. Regarding both symmetries mentioned above, a few comments are in order:
\begin{itemize}
\item The magic symmetry is found to be naturally related to the tri-bimaximal (TBM) pattern of mixing \cite{Lam:2006wy}. This can also be achieved by modified versions of the TBM mixing matrix, such as $TM_2$, to account for non-zero reactor angle and $CP$ violation \cite{Channey:2018cfj,Ouahid:2023hzs,Ding:2024ozt}. Such a patterns are related to a $\mu-\tau$ permutation symmetry rather than a reflection symmetry between the mass matrix entries, implying, in consequence, different symmetry transformations between neutrino states. 
\item The mixing pattern compatible with an exact magic symmetry presents tight relations between the mixings and the Dirac $CP$-violating phase \cite{Verma:2019uiu,Singh:2022nmk}. Majorana phases can be included within this approach and related to the mixing angles using two free parameters. Such parameters lead to sharp predictions for the Dirac phase, atmospheric, and solar angles, given the correlations to the reactor angle. Still, Majorana phases spread out over a wider range. 
\item Previous works have studied combinations of $TM_{1,2}$ mixings and $\mu-\tau$ reflection symmetry both at the Majorana neutrino mass matrix level and complete seesaw level \cite{Rodejohann:2017lre,Zhao:2021dwc}. Such a framework is highly restrictive and predictive, where the reactor angle is the only free parameter. It is worth noticing, within these approaches, that the obtained mass matrix does not necessarily possess a magic symmetry. 
\item Although there is no experimental evidence suggesting a magic symmetry, it is of phenomenological interest to scan the possibility of being present with other neutrino symmetries, which could motivate deeper studies. Previous works have shown that the Majorana phases of a cobimaximal mixing break the $\mu-\tau$ reflection symmetry \cite{Rivera-Agudelo:2022qpa}, giving a new approach to handle the symmetry breaking, with potential implications on the magic symmetry. 
\end{itemize} 

In the following, we adopt a \textit{naive}, model-independent framework with the principal aim of exploring potential imprints of a magic symmetry within the context of an approximately realized reflection symmetry in the neutrino mass matrix. The distinctive feature of our approach is the choice of the cobimaximal mixing pattern as the initial ansatz, rather than a $TM_2$-type structure, while refraining from introducing any correction parameters beyond the Majorana phases. This allows for a simultaneous investigation of possible manifestations of magic symmetry.

\section{Deviations from a symmetric scenario}
\label{sec:deviations}

It has been previously noted that including Majorana phases in a CBM matrix causes the breaking of the $\mu-\tau$ reflection symmetry in the mass matrix. This breaking can be adjusted by expressing the mass matrix in the following form \cite{Rivera-Agudelo:2022qpa,Xing:2015fdg,Rivera-Agudelo:2020orh}: 
\begin{equation}
\label{eq:corrected}
M_{\nu}= M_{\nu}^{0} + \delta M ,
\end{equation}
where $M_{\nu}^{0}$ denotes a symmetric matrix as in Eq.~(\ref{eq:massmatrix}), and
%
%e6 #&#
\begin{equation}
\label{eq:corrmass}
\delta M  = \left(
\begin{array}{c@{\quad}c@{\quad}c}
\epsilon_1 & 0 & \delta_2
\\
0 & 0 & \delta_1
\\
\delta_2& \delta_1 & \epsilon_2
\end{array}
\right)
\end{equation}
modulates the deviations from reflection symmetry by means of the four breaking parameters  $\delta_1=Im(m_{\mu\tau})$, $\delta_2= m_{e\tau}-m_{e\mu}^{*}$, $\epsilon_1 = Im(m_{ee})$ and $\epsilon_2=m_{\tau\tau} - m_{\mu\mu}^{*}$. Also, to avoid cases where the mass matrix entries are tiny by themselves, but do not respect the symmetric limit, we can define the dimensionless breaking parameters in the following way
\begin{eqnarray}	\label{eq:deltaepsilon}
	\hat{\delta}_1 &\equiv &\frac{\delta_1}{Re(m_{\mu\tau})} = \nonumber \frac{c_{12}^2 (m_2 \sin 2\beta_2 + m_1 \sin 2\beta_1 s^2_{13} )+s^2_{12}(m_1\sin 2\beta_1 + m_2 \sin 2\beta_2 s^2_{13})}{m_1\cos 2\beta_1(s^2_{12}+c^2_{12}s^2_{13})+m_2\cos 2\beta_2 (c^2_{12}+s^2_{12}s^2_{13})-m_3c^2_{13}} ~, \nonumber
	\\
	\hat{\epsilon}_1 &\equiv &\frac{\epsilon_1}{Re(m_{ee})} = \nonumber \frac{c^2_{13}(m_1\sin 2\beta_1 c^2_{12} + m_2 \sin 2\beta_2 s^2_{12})}{c^2_{13}(m_1 \cos 2\beta_1 c^2_{12} + m_2 \cos 2\beta_2 s^2_{12})-m_3s^2_{13}} ~,
\nonumber
\\
	\hat{\delta}_2 &\equiv &\frac{\delta_2}{m_{e\mu}^{*}} = 1-\frac{i m_3 s_{13}-e^{2i\beta_1}m_1 c_{12}(s_{12}-ic_{12}s_{13})+e^{2i\beta_2}m_2 s_{12}(c_{12}+is_{12}s_{13})}{i m_3 s_{13}-e^{-2i\beta_1}m_1 c_{12}(s_{12}-ic_{12}s_{13})+e^{-2i\beta_2}m_2 c_{12}(c_{12}+is_{12}s_{13})}~,
	\nonumber
	\\
	\hat{\epsilon}_2 &\equiv & \frac{\epsilon_2}{m_{\mu\mu}^{*}} =
	1-\frac{i m_3 c_{13}^2+e^{2i\beta_1}m_1(s_{12}-ic_{12}s_{13})^2+e^{2i\beta_2}m_2(c_{12}+is_{12}s_{13})^2}{i m_3 c_{13}^2+e^{-2i\beta_1}m_1(s_{12}-ic_{12}s_{13})^2+e^{-2i\beta_2}m_2(c_{12}+is_{12}s_{13})^2}~. 
\end{eqnarray}
After some algebra, it is direct to obtain expressions for these parameters as shown in the right-hand side of Eq. (\ref{eq:deltaepsilon}), where, it is clear to see that the smallness of these parameters is related to the Majorana phases. As has been previously noted, Majorana phases play an important role in determining the size of the deviations from $\mu-\tau$ reflection symmetry \cite{Rivera-Agudelo:2022qpa}. 

%%%%%%%%%%%%%%%%%%%%%%%%%%%%%%%%%%%%%%%
%%%%%%%%%%%%%%%%%%%%%%%%%%%%%%%%%%%%%%%%%%%%%%%

We can easily verify from Eq. (\ref{eq:deltaepsilon}) that an exact $\mu-\tau$ reflection symmetry is recovered for any combination of Majorana phases with values $0, \pm \pi/2,\pm \pi$ \cite{Xing:2022uax}. It is important to note that, in addition to the trivial case of zero phases, there are possible combinations where one phase, or both, may maximally violate the $CP$ symmetry. One might expect that the Majorana phases would remain close to the previous combinations for small deviations from the symmetric limit. However, it is crucial to highlight that different mass orderings need to be considered. As indicated by the expressions in Eq. (\ref{eq:deltaepsilon}), in the case of normal ordering (where \( m_1 \ll m_2 < m_3 \)), the breaking parameters do not show a strong dependence on \( \beta_1 \), which leaves this phase undetermined. This phase could be constrained by considering additional restrictions from magic symmetry. It is worth mentioning that the Majorana phases as included in Eq. (\ref{eq:corrected}) only affect the mass matrix elements, and hence the breaking parameters, but not modify the value of the mixing angles nor the Dirac phase\footnote{Previous works have explored the possibility of correcting the cobimaximal mixing to account for values of the atmospheric angle and the Dirac phase far from maximality \cite{Gomez-Izquierdo:2023mph}, which usually arises from corrections to the charged lepton sector, but this is out of the scope of the present work.}. 

On the other hand, if we demand the neutrino mass matrix to be magic, we have only to pay attention to the sum of each row (or column) because of the symmetry property. Then, we can fully compute these sums as $S_{\ell} = \sum_{i=1}^3 [M_{\nu}]_{\ell i}$, with $\ell = e,\mu,\tau$, for the first, second and third row, respectively. We can search for deviations between these results to explore the contributions of $CP$ phases. By the same arguments of Eq. (\ref{eq:deltaepsilon}), we propose to modulate the deviations from magic symmetry by defining two dimensionless parameters

\begin{eqnarray}
\label{sumas}
\hat{S}_{\mu} &\equiv& \frac{S_{\mu}-S_e}{S_{\mu}} ~,\nonumber \\
\hat{S}_{\tau} &\equiv& \frac{S_{\tau}-S_e}{S_{\tau}} ~,
\end{eqnarray}
where we have chosen to consider deviations with respect the first sum ($S_e$), where the neutrinoless double beta decay amplitude $m_{\beta \beta}= [M_{\nu}]_{1 1}$ is of great interest. After some algebra, it is possible to express the breaking parameters of the magic symmetry as
\begin{eqnarray}
	\label{eq:magicbreak}
	\hat{S}_{\mu} &=& 1-\frac{2\tilde{m_1}(a-c_{13}^2)-2a\tilde{m_2}+2m_3 s_{13}^2}{\tilde{m_1}(b-is_{13}(c-\sqrt{2}c_{13}))-\tilde{m_2}(2+b-ics_{13})+i\sqrt{2}m_3 c_{13}s_{13}} \nonumber \\
	\hat{S}_{\tau} &=&  1-\frac{2\tilde{m_1}(a-c_{13}^2)-2a\tilde{m_2}+2m_3 s_{13}^2}{\tilde{m_1}(b+is_{13}(c-\sqrt{2}c_{13}))-\tilde{m_2}(2+b+ics_{13})-i\sqrt{2}m_3 c_{13}s_{13}} ~,
\end{eqnarray}
in a similar fashion to Eq. (\ref{eq:deltaepsilon}). We have now introduced the compact notation $\tilde{m_j}=m_j e^{2i\beta_j}$, for $j=1,2$, and
\begin{eqnarray}
	a &=& s_{12}^2 c_{13}^2 + \sqrt{2} s_{12} c_{12} c_{13} \nonumber\\
	b &=& \sqrt{2} s_{12}c_{12}c_{13} - 2s_{12}^2 \\
	c &=& 2s_{12} c_{12} +\sqrt{2} s_{12}^2 c_{13} ~ \nonumber .
\end{eqnarray}
The dependence of Eq. (\ref{eq:magicbreak}) on mixing, masses, and phases is clear. Therefore, we can anticipate that Majorana phases may be constrained in scenarios involving small deviations, especially considering the current precision of the mixing angles. Let us explore some specific cases that can be identified from these expressions.

It is direct to see that the difference between both expressions in Eq. (\ref{eq:magicbreak}) is due to the imaginary term, which is also proportional to $s_{13}$. Given the smallness of the reactor angle, we can neglect it in a first approximation. In such a case, we are left with identical breaking parameters ($\hat{S}_{\mu}=\hat{S}_{\tau}$), which are proportional to $(\tilde{m_1}-\tilde{m_2})(2s_{12}+\frac{1}{\sqrt{2}}s_{12}c_{12}-1)$. Hence, the exact symmetry condition would demand this product to be zero, which could be fulfilled in two ways. On the one hand, regarding the solar angle, we can verify that the selection $s_{12}=1/\sqrt{3}$ ($c_{12}=\sqrt{2/3}$) would respect the magic symmetry, independently of the mass ordering and Majorana phases. On the other hand, the equality $\tilde{m_1}=\tilde{m_2}$ would also give a symmetric scenario, which is limited by the equality in neutrino masses and Majorana phases ($\beta_1=\beta_2$). We can identify the former case as the imprints of the tribimaximal mixing pattern, consistent with previous studies of magic symmetry, and the last with the preference for a certain combination of Majorana phases. However, both exact scenarios are disfavoured by experimental determinations given that $\theta_{13}\neq 0$.

We have shown that each exact symmetry requirement on the neutrino mass matrix may lead, by itself, to specific combinations of Majorana phases and/or the mixing angles. Eqs. (\ref{eq:deltaepsilon}) and (\ref{eq:magicbreak}) show that the six breaking parameters shall depend only on the Majorana phases, and the lightest neutrino mass, when the mixing angles are fixed to the experimental values. While the lightest neutrino mass can be restricted from the experimental bounds, relevant data for Majorana phases is absent. Hence, the breaking parameters of both symmetries are correlated only through the Majorana phases, once we choose a value for $m_0$. Next, we can adopt an alternative approach that consists of limiting the values of the breaking parameters rather than the phases. We propose to impose small deviations from both (magic and reflection) symmetric patterns and demand that the breaking parameters in Eqs. (\ref{eq:deltaepsilon}) and (\ref{eq:magicbreak}) to be small,  which is also called a light or soft breaking of the symmetries \cite{Xing:2015fdg,Xing:2022uax}. Slightly relaxing these values from the exact symmetry might help narrow down the CP phases consistent with both sets of conditions, in the two mass orderings. In the following, let us perform a numerical scan of the proposed scenario and discuss our results.

%%%%%%%%%%%%%%%%%%%%%%%%%%%%%%%%%%%%%%%%%%%%%%%%%%%%%%%%%%%%%%%%%%%%%%
%%%%%%%%%%%%%%%%%%%%%%%%%%%%%%%%%%%%%%%%%%%%%%%%%%%%%%%%%%%%%%%%%%%%%%

%s3 #&#
\section{Results}
%%LEAP%%%\label{sec3}
%%LEAP%%%\label{sec4}
\label{sec:results}

Our previous semi-analytical treatment will serve as a guide in our numerical search for studying both symmetric limits. As we could see, it will be relevant to consider both (normal and inverted) mass orderings with the corresponding experimental results for the mixings. In Tab. (\ref{tab:values}), we show the experimental values considered in our numerical\footnote{Similar results are obtained for other data sets.} evaluations \cite{Esteban:2020cvm}, while the atmospheric angle and Dirac phase are fixed to cobimaximal values. In the normal ordering, $m_1=m_0$ is considered the lightest neutrino mass, while $m_2^2=m_0^2 + \Delta m_{21}^2$ and $m_3^2=m_0^2 + \Delta m_{3\ell}^2$, where $\Delta m_{21}^2$ and $\Delta m_{3\ell}^2$ are the solar and atmospheric squared mass differences, respectively. For the inverted ordering, $m_3=m_0$ is taken as the lightest mass, and $m_1^2=m_0^2 - \Delta m_{3\ell}^2$ and $m_2^2=m_0^2 -  \Delta m_{3\ell}^2  + \Delta m_{21}^2$.

\begin{table}[t]
\caption{Experimental values considered in the analysis as reported in \cite{Esteban:2020cvm}. Three sigma intervals are presented for mixings and one sigma uncertainties for the squared mass differences.}\label{tab:values}
\begin{tabular}{@{}lll@{}}	
		\toprule
		 & Normal Ordering &  Inverted Ordering \\ 
		\midrule
		 $\sin^2 \theta_{12}$ & $0.275 \rightarrow 0.344$ & $0.275 \rightarrow 0.344$ \\

		 $\sin^2 \theta_{13}$ & $0.02047 \rightarrow 0.02397$ & $0.02049 \rightarrow 0.02420$ \\

		  $\frac{\Delta m_{21}^2}{10^{-5}\text{eV}^2}$ & $7.41^{+0.21}_{-0.20}$ & $7.41^{+0.21}_{-0.20}$ \\

		  $\frac{\Delta m_{3\ell}^2}{10^{-3}\text{eV}^2}$ & $+2.505^{+0.024}_{-0.026}$ & $-2.487^{+0.027}_{-0.024}$ \\
	\botrule
\end{tabular}
\end{table}

To explore small deviations from magic and reflection symmetry, we will consider the three-sigma ($3\sigma$) range of the mixing angles. At the same time, for the sake of simplicity, we will keep the central values of squared mass differences\footnote{We have verified that the breaking parameters have no visible changes when the $3\sigma$ range is considered for the squared mass differences. On the contrary, they have shown strong dependence on the chosen value for $m_0$ in each mass ordering.}. Hence, depending on the mass hierarchy, we are left with three nonfixed parameters: $m_0$ (the lightest neutrino mass), and the Majorana phases ($\beta_1$ and $\beta_2$).  $m_0$ will be considered up to the upper limit of $0.3$ eV, while Majorana phases will be scattered in the full ($-\pi,\pi$) range. Hence, we will explore two cases, one where the breaking parameters are at most 0.1, and, for comparison, one where they are relaxed to be at most 0.3. Values greater than 0.3 could be related to large breaking and not be considered in our discussion.  

%f1 #&#
\begin{figure}
\includegraphics[width=0.45\textwidth]{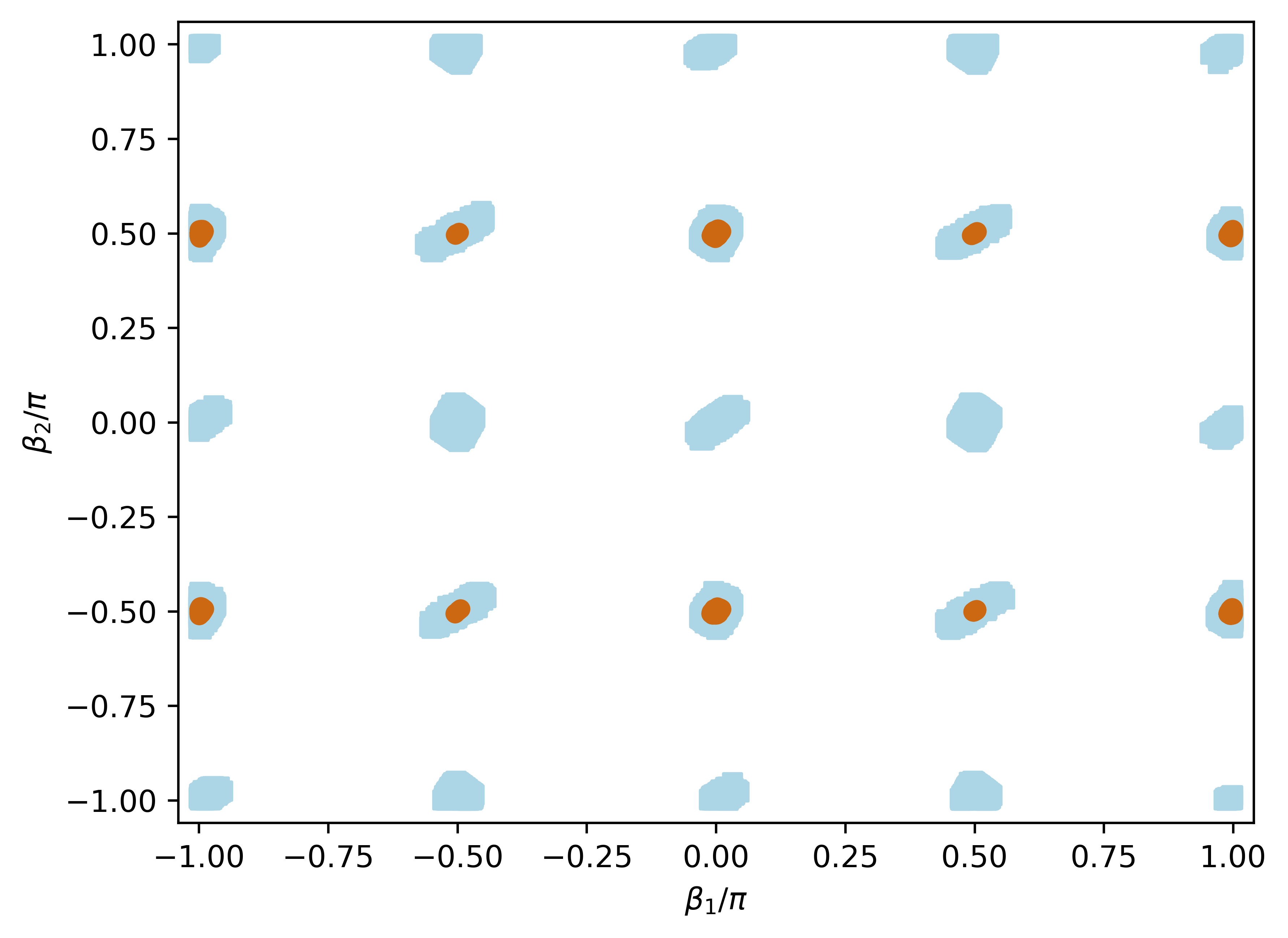} 
\includegraphics[width=0.45\textwidth]{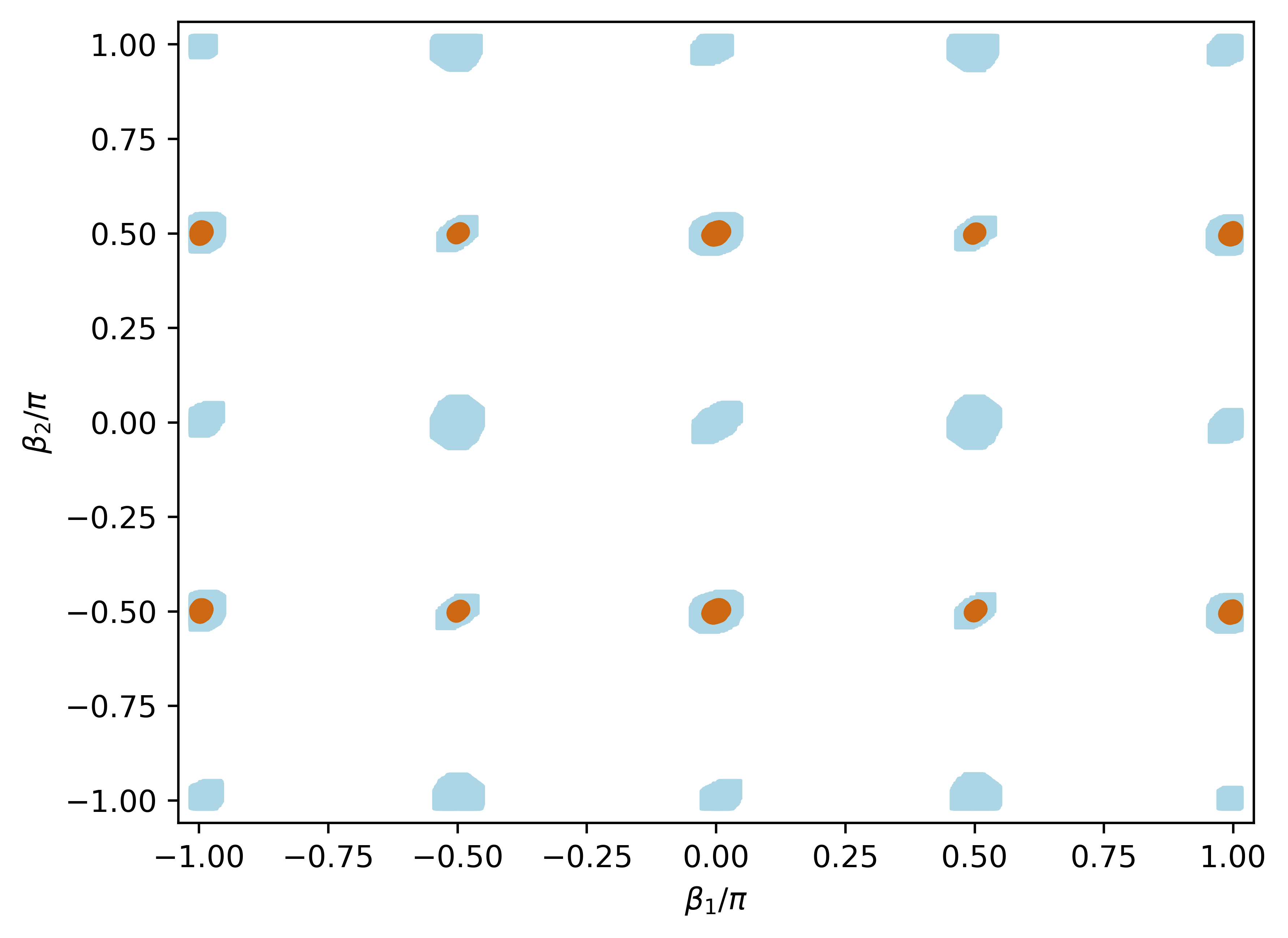}
\caption{Allowed regions of Majorana phases for $NO$ (left) and $IO$ (right).
Small (orange)  and large (light blue) regions correspond to Max$[|
\hat\delta_i| , |\hat\epsilon_i|,|\hat S_{\mu}|,|\hat S_{\tau}|] \lesssim 0.1,$ and $~0.3$, respectively.}%
\label{fig:betas}
\vspace*{-4pt}
\end{figure}

We present in Fig. \ref{fig:betas} the allowed values of Majorana phases consistent with small departures from the symmetries in both mass orderings. We can observe that these regions, for values of breaking parameters up to $0.3$ (Max$[|\hat\delta_i| , |\hat\epsilon_i|,|\hat S_{\mu}|,|\hat S_{\tau}|] \lesssim 0.3$, with $i=1,2$), remain near the symmetry conserving combinations identified in our previous analysis. We recognize that these restrictions (light blue regions) are mainly dominated by the reflection symmetry parameters, which shows that the condition of equal Majorana phases, arising from magic symmetry, is relaxed in this case and is compensated by the values of the reactor and solar angles. We can also observe slight differences between both mass orderings for certain combinations of these phases, which justify our attention to lightest neutrino mass contributions. Nevertheless, regions are visibly reduced as we reach the symmetric limit (Max$[|\hat\delta_i| , |\hat\epsilon_i|,|\hat S_{\mu}|,|\hat S_{\tau}|]  \lesssim 0.1$, orange regions). As was previously anticipated, only a few combinations of Majorana phases could be consistent with a small symmetry breaking, which  we identify as near the following values: $|\beta_1| \approx 0,\pi/2,\pi$, and $|\beta_2| \approx \pi/2$. Such a combinations are the result of considering the full expressions of breaking parameters and not only the approximated cases, which also include contributions of non zero $\theta_{13}$. It is interesting to note that some $CP$ conserving combinations of Majorana phases are excluded within the very small symmetry breaking requirement, as is the case of zero Majorana phases. 

\begin{figure}
\includegraphics[width=0.45\textwidth]{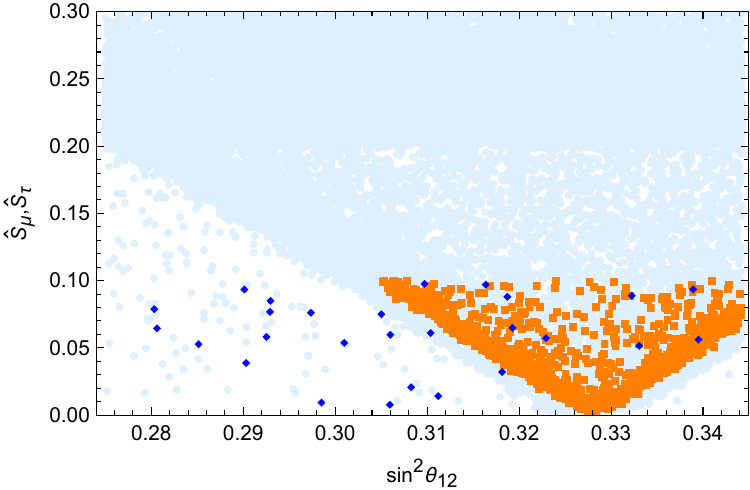} 
\includegraphics[width=0.45\textwidth]{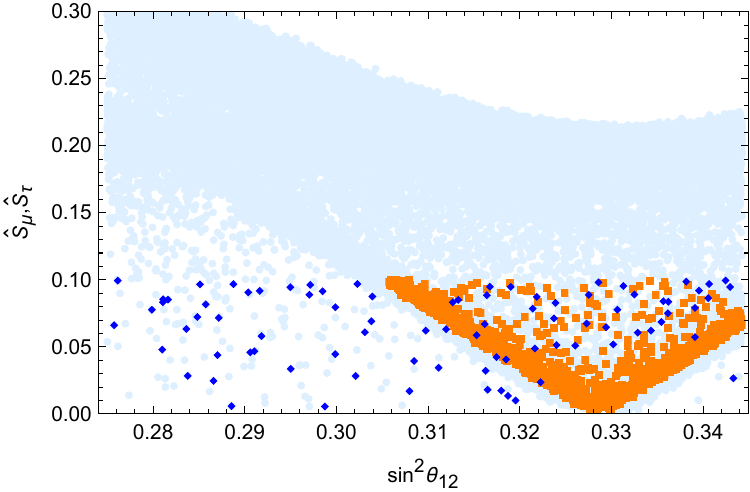}
\caption{Deviation parameters from magic symmetry for different values of $\sin^2\theta_{12}$ in normal (left plot) and inverted (right plot) orderings. The orange (squares) and light blue (circles) scattered points correspond to Max$[|\hat\delta_i| , |\hat\epsilon_i|,|\hat S_{\mu}|,|\hat S_{\tau}|] \lesssim 0.1, ~0.3$, respectively. Blue diamonds represent the case of $|\beta_{1,2}|\approx \pi/2$ for very small breaking parameters ($\lesssim 0.1$).}%
\label{fig:svs12}
\vspace*{-4pt}
\end{figure}

The smallness condition of breaking parameters seems to point, at the end, to specific values of mixings and phases. We show in Fig. (\ref{fig:svs12}) the plots related to the breaking parameters of magic symmetry compared with the solar angle. As mentioned in our previous discussion, the smallness of magic symmetry parameters points towards a trimaximal value for the solar angle ($\sin^2 \theta_{12}\sim 1/3$) when different modulus of Majorana phases are considered ($|\beta_1| \neq |\beta_2|$). On the other hand, it is also possible to lay far from this value when cancelations in $(\tilde{m_1}-\tilde{m_2})$ occur. This is shown in Fig. (\ref{fig:svs12}) by isolating the points corresponding to $|\beta_1| \approx |\beta_2| \approx \pi/2$.

%f2 #&#

\begin{figure}\centering
	\includegraphics[width=0.5\textwidth]{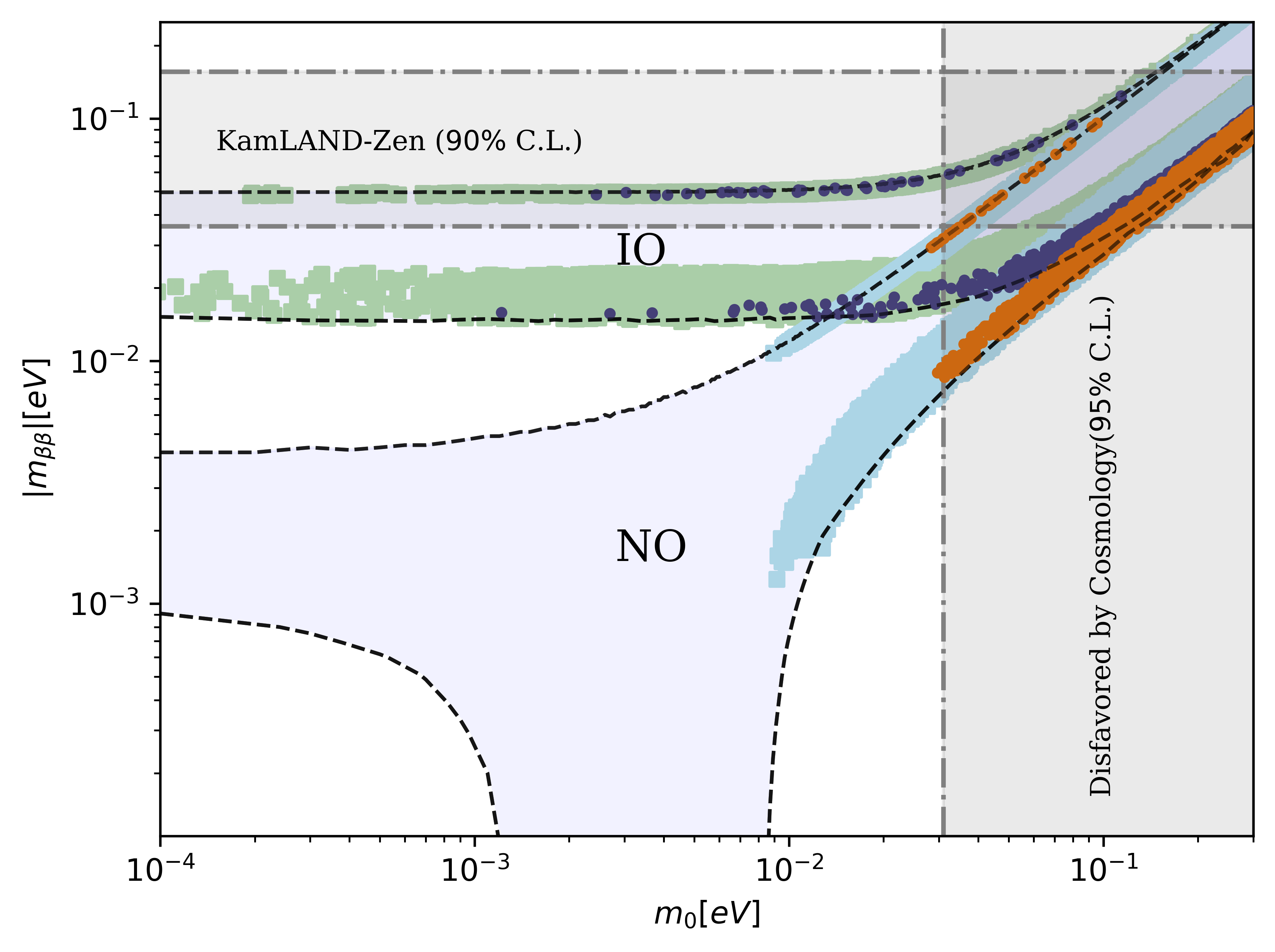} 
\caption{Allowed regions of $|m_{\beta\beta}|$ for different lightest neutrino mass in the $NO$ and $IO$. Light blue and green squares correspond to Max$[|\hat\delta_i| , |\hat\epsilon_i|,|\hat S_{\mu}|,|\hat S_{\tau}|] \lesssim 0.3$, and purple and orange points to very small breaking parameters ($\lesssim 0.1$).  Regions delimited by dotted lines represent the full region of $NO$ and $IO$ for nonrestricted $CP$ phases. Black star marks the best-fit point of Majorana phases in both orderings for $m_0=0.02 eV$. Gray shaded areas show the excluded regions by cosmology and $0\nu\beta\beta$ experiments \cite{KamLAND-Zen:2022tow,Planck:2018vyg,DESI:2024mwx}.}% 
\label{fig:doublebeta}
\vspace*{-4pt}
\end{figure}

To confront the restrictions in $CP$ phases with experimental observables, we show in Fig. \ref{fig:doublebeta} the corresponding regions for the neutrinoless double beta decay amplitude $|m_{\beta \beta}|$ as a function of the lightest neutrino mass. We also show the current experimental bounds from neutrinoless double beta decay and cosmology collaborations \cite{KamLAND-Zen:2022tow,Planck:2018vyg,DESI:2024mwx}. The regions corresponding to unrestricted phases are also shown and can be compared to the restricted regions arising from considerations of small deviations. The obtained regions may serve to test the possibility of having a slightly broken scenario with the results of forthcoming experiments. 
 
It is worth noticing that constraints in neutrinoless double beta decay amplitude from the most recent analysis of KamLan-Zen Collaboration \cite{KamLAND-Zen:2022tow} rule out the upper band of IO in Fig. \ref{fig:doublebeta}, which corresponds to the combinations $(|\beta_1|,|\beta_2|)\approx(0,0)$, $(0,\pi)$, $(\pi,0)$, $(\pi/2,\pi/2)$, $(\pi,\pi)$. Hence, only the lower band is currently allowed, which is identified with the combinations of different phases $(|\beta_1|,|\beta_2|)\approx (0,\pi/2)$, $(\pi/2,0)$, $(\pi,\pi/2)$, $(\pi/2,\pi)$. From the right plot in Fig. \ref{fig:betas}, in addition to these constraints, we obtain that values corresponding to $(|\beta_1|,|\beta_2|)\approx(\pi,\pi/2),(0,\pi/2)$ are the only combinations allowed for breaking parameters $\lesssim 0.1$, which are within the most restricted region of the solar angle in Fig. \ref{fig:svs12} (orange squares).  
 
%f3 #&#

\begin{table}[h]
\caption{Best fit values of Majorana phases for different $m_0$. Predicted $m_{ee}$ and the largest breaking parameter (LBP) in each case are also shown.}\label{tab:majoranas}
\begin{tabular}{@{}lll@{}}	
		\toprule
		 & Normal Ordering &  Inverted Ordering \\ 
	\midrule
		$m_0 = 0.1 ~eV$ & & \\
		 $(\beta_1,\beta_2)$ & $(1.57,3.06)$ & $(1.58,3.08)$ \\

		 $|m_{\beta\beta}|$ & $0.034$ & $0.045$ \\

		 LBP & $|\hat S_{\tau}|=0.24$ &  $|\hat S_{\tau}|=0.23$ \\
	\midrule
		$m_0 = 0.02 ~eV$ & & \\
		 $(\beta_1,\beta_2)$ & $(1.64,-1.50)$ & $(0.06,0.05)$ \\

		 $|m_{\beta\beta}|$ & $0.021$ & $0.052$ \\

		 LBP & $|\hat\epsilon_2|=0.17$ & $|\hat\delta_1|=0.16$ \\
	\midrule
		$m_0 = 0.001 ~eV$ & & \\
		 $(\beta_1,\beta_2)$ & $(1.39,-1.65)$ & $(1.62,3.14)$ \\

		 $|m_{\beta\beta}|$ & $0.005$ &  $0.019$\\

		 LBP & $|\hat S_\mu|=0.53$ &  $|\hat\epsilon_1|=0.20$ \\
		 \midrule
\end{tabular}
\end{table}

Restriction on the lightest neutrino mass coming from cosmology \cite{Planck:2018vyg,DESI:2024mwx} is also shown in Fig. \ref{fig:doublebeta} for comparison.  We can obtain the best-fit values of Majorana phases considering the current bounds. As an example, we can minimize the breaking parameters within the range $(0,0.3)$,  by taking the central values in Table \ref{tab:values}, and different values of $m_0$. We show in Tab. (\ref{tab:majoranas}) the best fit values of the Majorana phases for the three values of the lightest neutrino mass, $0.1~eV$, $0.02~eV$, and $0.001~ eV$, which are above, near, and far from the current bound, respectively. We can identify that the small breaking scenario of magic and cobimaximal symmetries favours certain combination of CP phases which may differ between the mass orderings, and depends of the lightest neutrino mass selection. We observe that for the degenerate region ($m_0=0.1~eV$), both mass orderings coincide in favour of $CP$ violation for $\beta_1$ and near $CP$ conservation for $\beta_2$. While this scenario is compatible with small symmetry breaking, where the largest breaking parameter (LBP) is around $0.2$, is excluded by Cosmology results. For an allowed value near the current limit ($m_0=0.02~eV$), where the LBP is also around $0.2$, $CP$ violation is preferred in the normal ordering and $CP$ conservation for the inverted ordering. Finally, for $m_0=0.001 ~eV$, we observe that $CP$ violation is preferred by the normal ordering. However, such a scenario is incompatible with the small symmetry breaking requirement in the magic symmetry as the LBP is of order $0.5$, consistent with Fig. (\ref{fig:doublebeta}). Nevertheless, the symmetry requirements can be fulfilled in the inverted ordering, favouring $CP$ violation in $\beta_1$ and $CP$ conservation in $\beta_2$.

The most restrictive sensitivity from nEXO (with $|m_{\beta\beta}| \approx 10$ m$eV$) \cite{nEXO:2021ujk} would rule out the inverted ordering (IO) and the upper light blue region of the normal ordering (NO) in Fig. \ref{fig:doublebeta}. In this scenario, only the combinations $(|\beta_1|, |\beta_2|) \approx (\pi, \pi/2)$ and $(0, \pi/2)$, along with a lightest neutrino mass above $0.008$ eV, would be allowed under the small breaking condition. This corresponds to the lower region of the NO (indicated by the light blue squares). Consequently, this sets the stage for either confirming or dismissing the scenario of small deviations in the near future.

%s4 #&#
\section{Summary}
%%LEAP%%%\label{sec4}
\label{sec:conclusions}
\vspace*{-3pt}

 The searches for symmetries in the neutrino matrices could help to elucidate the current pattern of masses and mixings. They would serve as a guide in the search of a more fundamental flavor theory. From a model-independent point of view, we have explored the possibility of having a magic neutrino mass matrix consistent with the reflection symmetry, where the atmospheric angle and Dirac $CP$ phase are fixed to cobimaximal values.

It was found that, in both mass hierarchies, it is possible to accommodate a slightly broken scenario rather than an exact one, given the experimental restrictions of mixing angles. Such considerations are consistent with current global fits and were used to restrict the values of Majorana phases to specific combinations, which were further reduced near the symmetric limit. Also, there was a preference for the trimaximal value of the solar angle, in the case of non-equal Majorana phases, leading, in consequence, to a particular pattern of mixings, where only the reactor angle would remain free. The former considerations gave rise to tight regions in the double beta decay amplitude, which were faced with experimental bounds. Forthcoming results of neutrinoless double beta decay could test the IO and a part of the NO. 

Both well-known symmetries were explored simultaneously for the first time, which may hold theoretical interest. Our approach has shown that, in the NO, only neutrino masses above $0.008$ eV would be consistent with the small breaking requirement, favouring $CP$ violation in both phases. On the other hand, in the IO, the lightest mass could take very small values. The projected sensitivities on $m_{\beta\beta}$ may test this region, allowing us to determine the size of the breaking and the combinations of CP phases.

%%%%%%%%%%%%%%%%%%%%%%%%%%%%%%%%%%%%%%%%%%%%%%%%%%%%%%%%%%%%%%%%%%%%%%%%%%%%%%%%%%%%%%%%
 %%%%%%%%%%%%%%%%%%%%%%%%%%%%%%%%%%%%%%%%%%%%%%%%%%%%%%%%%%%%%%%%%%%%%%%%%%%%%%%%%%%%%%%%%%%%%% 
   
\section*{Acknowledgements}

 The authors acknowledge funding from  \textit{Division General de Investigaciones} (DGI) of the Santiago de Cali University under grant 935-621124-655. D. C. R. thanks Minciencias under grant CD-82315 CT ICETEX 2021-1080.

\bibliographystyle{utphys}
\vspace*{-10pt}
%\bibliography{Bibliography.bib}
%%%%%%%%%%%%%%%%%%%%%%%%%%%%%%%%%%%%%%%%%%%%%%%%%%%%%%%%%%%%%%%%%%%%%%%%%%%%%%%%%%%%%%%%%%%%%

\providecommand{\href}[2]{#2}\begingroup\raggedright\endgroup

%%%%%%%%%%%%%%%%%%%%%%%%%%%%%%%%%%%%%%%%%%%%%%%%%%%%%%%%%%%%%%%%%%%%%%%%%%%%%%%%%%%%%%%%%%%
%%%%%%%%%%%%%%%%%%%%%%%%%%%%%%%%%%%%%%%%%%%%%%%%%%%%%%%%%%%%%%%%%%%%%%%%%%%%%%%%%%%%%%%%%%%

\end{document}